\documentclass[twocolumn,twoside,slac_two]{revtex4}
\usepackage{graphicx}
\usepackage{fancyhdr}
\def\D0{D\O}  \def\d0{D\O}

\begin{document}

\title{D\O\ Hot Topics}

%

\author{D. Buchholz on behalf of the D\O\ Collaboration}
\affiliation{Northwestern University, Department of Physics and Astronomy, Evanston, Il 60208, USA}
\begin{abstract}
We present recent D\O\ results based on approximately $1$ fb$^{-1}$ of $p \bar{p}$
collisions at $\sqrt{s} = 1.96$ TeV recorded at the Fermilab Tevatron.
Preliminary results on a search for the flavor changing neutral current
 process $D^+\rightarrow \pi^+\mu^+\mu^-$, a measurement of the CP violation parameter
 in $B$ mixing, $\epsilon_B$,  and a two sided limit on the $B_s$
 oscillation frequency $\Delta m_s$ are presented.  The limits on $\epsilon_B$ and
 ${\cal B}(D^+\rightarrow \pi^+\mu^+\mu^-)$ are the world's best limits. 
 The two sided bound on $\Delta m_s$ is the first direct indication by a
 single experiment that $\Delta m_s$ is bounded from above.
\end{abstract}

\maketitle

\thispagestyle{fancy}

\section{Introduction}
A major goal of the heavy flavor physics community is to find indirect evidence of new physics
 by studying neutral meson mixing and flavor changing neutral current (FCNC) decays.  In the Standard
 Model (SM) these processes are mediated by higher order weak transitions and can receive sizeable
 corrections from new particles as predicted by several extensions to the SM. These new physics
 corrections are manifested in either an augmentation of the decay or mixing rate, or possibly
 by introducing new complex operators leading to unexpected sources of CP violation.

 The D\O\ experiment operating at the Fermilab Tevatron has a rich heavy flavor program built on the
 excellent performance of the D\O\ muon system. The large single and dimuon data samples are ideal for
 studying heavy flavor mixing and radiative decay. In this presentation, we focus on three results that
 highlight the strengths of the program:  a search for the flavor changing neutral current transition
 $c\rightarrow u \mu^+\mu^-$~\cite{bib:conf1}, a search for CP violation in $B$ mixing~\cite{bib:conf2}, and our progress in measuring
 the $B_s$ mixing frequency~\cite{bib:prl}.  These results are derived from a data sample of  $p\bar{p}$ collisions at
 $\sqrt{s}=1.96$ TeV corresponding to a luminosity of 
approximately $1$ fb$^{-1}$.

\section{ D\O\ experiment}

The D\O\ detector is a general purpose spectrometer
and calorimeter~\cite{bib:dzero}.
Charged particles are reconstructed using a silicon vertex tracker
 and a scintillating fiber tracker located inside a
superconducting solenoid coil that provides a $2$ T magnetic field.
Photons and electrons are reconstructed using the inner region of a liquid argon
calorimeter optimized for electromagnetic shower detection. Jet
reconstruction and electron identification are further augmented with
the outer region of the calorimeter optimized for hadronic shower
detection. 
Muons are reconstructed using a
spectrometer consisting of magnetized iron toroids and three super-layers of
proportional tubes and plastic trigger scintillators located outside the calorimeter. 

The D\O\ trigger is based on a three tier system.  The level 1 and 2
dimuon triggers rely on energy deposited in the muon spectrometer and
fast reconstruction of muon tracks.  For single muon triggers the muon is required to be
 associated with a track in the central detector with a transverse momentum of at least 
$3$ GeV/$c$. The level 3 trigger performs fast
reconstruction of the entire event allowing for further muon
identification algorithms, improved matching of muon candidates to tracks
reconstructed in the central tracking system, and requirements on the
$z$ position of the primary vertex.  At higher luminosities, we increase the penetration requirements
 on the dimuon triggers, the momentum thresholds are
 usually raised to as high as $5$ GeV/$c$ on the single muon triggers, and invariant mass or displaced particle
filters are added at level 3 to enhance the heavy flavor content of the muon samples.

\section{Flavor Changing Neutral Current Charm Decays}

The excellent agreement between observed FCNC processes involving down-type 
charge $-1/3$ quarks such as $b\rightarrow s \gamma$,
$b\rightarrow s l^+l^-$, and $K\rightarrow \pi \nu \bar{\nu}$ with SM
expectations have been used to set strict limits on new
phenomena~\cite{bib:kagen-neubert,bib:ali,bib:agashe}. 
 However, there are several scenarios of new phenomena such as SUSY R
parity violation in a single coupling scheme~\cite{bib:agashe}, or 
little Higgs models with a new up-like vector quark~\cite{bib:fajer2} 
where deviations from the SM would only be seen in the up sector. 
 Scenarios of this nature motivate the study of FCNC charm meson decays.

Due to GIM suppression, the SM rates for FCNC charm decays vanish in
the limit of SU(3) symmetry.  The inclusive rates of decays such as
$D^+\rightarrow \pi^+ l^+l^-$ are therefore expected to
be dominated by long distance contributions where the dilepton system
is produced via intermediate strong resonances such as the $\phi$ or the
$\omega$~\cite{bib:pakvasa,bib:fajer}.
Of particular interest is the decay $D^+_s\rightarrow \pi^+\mu^+\mu^-$ 
that can only proceed via the long distance interaction where the dimuon
 system is produced by an intermediate $\phi$ meson.  Since this has no 
short distance contributions, the rate is given simply by the product of
 the two branching fractions of $D^+_s\rightarrow \phi\pi^+$ and $\phi\rightarrow \mu^+\mu^-$.
 Just as the decay $B^+\rightarrow J/\psi
K^+\rightarrow K^+l^+l^-$ played a crucial role in benchmarking the
studies of $b\rightarrow sl^+l^-$ transitions, the observation of the Cabbibo favored decay
$D^+_s\rightarrow \phi\pi^+\rightarrow \pi^+l^+l^-$ is an essential
first step in the study of $c\rightarrow ul^+l^-$ transitions.
Once this mode has been observed, the search for continuum production
 of $c\rightarrow u l^+l^-$ can be performed by looking for an excess
 of $D^+$ candidates where the dilepton mass is inconsistent with resonance production.

$D$ meson candidates are formed by combining two well reconstructed muons with a
 track in the same jet as the dimuon system and requiring the three particle system
 have an invariant mass consistent with a $D$ meson.  If multiple candidates exist
 for a single event, the best candidate is selected based on the quality of the three
 track vertex, the transverse momentum of the pion, and the radial distance between
 the pion and the dimuon system.  Backgrounds from light quarks or random combinations
 are reduced further with variables that take advantage of the long $D$ meson lifetime 
and distinct decay topology.

\begin{figure}[h]
\centering
\includegraphics[width=80mm]{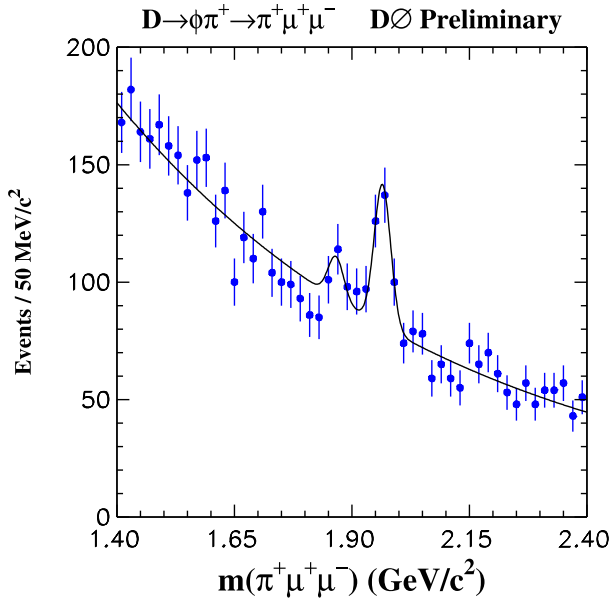}
\caption{The $m(\pi^+\mu^+\mu^-)$ mass spectrum for the loose cut sample in the $0.96 < m(\mu^+\mu^-) < 1.06$ GeV$/c^2$ bin.  The results of binned likelihood fits to the distributions
  including contributions for $D^+_s$, $D^+$, and combinatoric
  background are overlaid on the histogram.}
\label{fig:dres1}

\end{figure}

\begin{figure}[h]
\centering
\includegraphics[width=80mm]{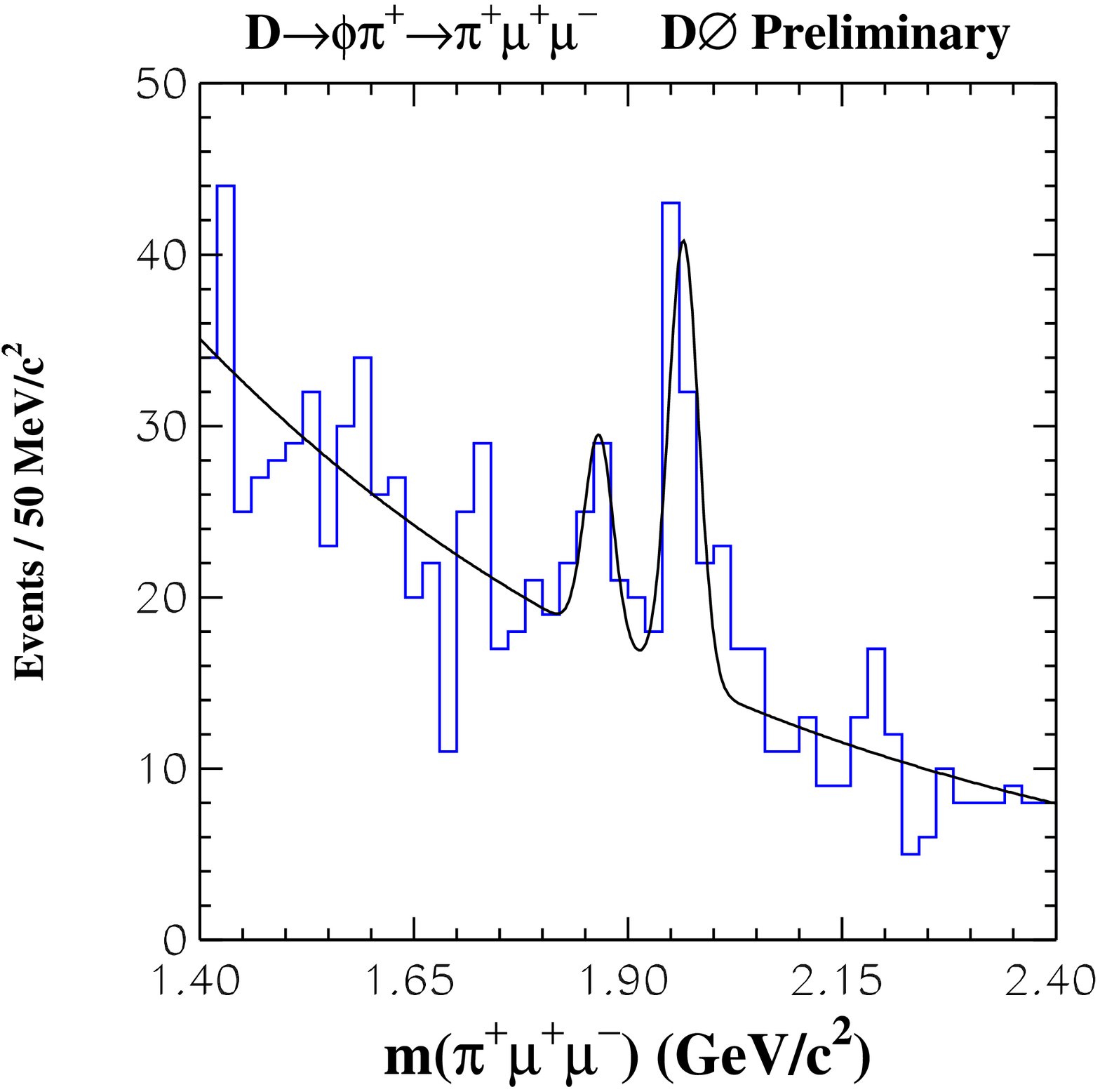}
\caption{The $m(\pi^+\mu^+\mu^-)$ mass spectrum for the optimized $D^+_s$ 
sample in the $0.96 < m(\mu^+\mu^-) < 1.06$ GeV$/c^2$ bin.  The results of 
binned likelihood fits to the distributions
  including contributions for $D^+_s$, $D^+$, and combinatoric
  background are overlaid on the histogram.}
\label{fig:dres1a}

\end{figure}
\begin{figure}[h]
\centering
\includegraphics[width=80mm]{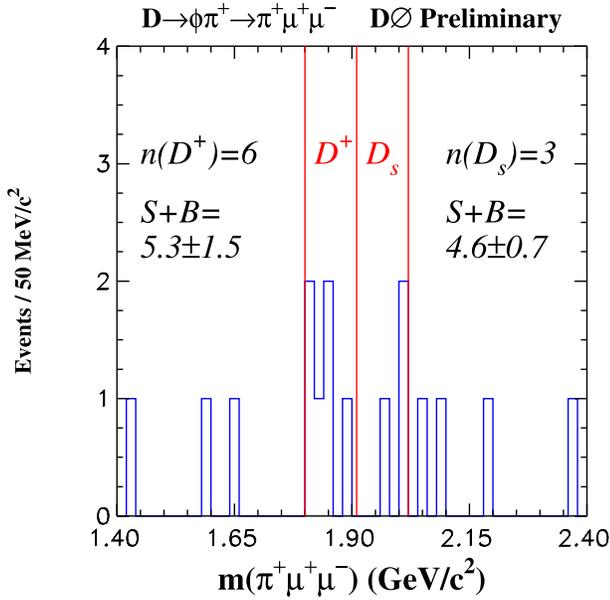}
\caption{The $m(\pi^+\mu^+\mu^-)$ mass spectrum for the optimized $D^+$ sample in the $0.96 < m(\mu^+\mu^-) < 1.06$ GeV$/c^2$ bin.  
The line is a fit to the data in the sideband regions $1.4 <  m(\pi^+\mu^+\mu^-)  < 1.7$ GeV$/c^2$ and 
 $2.1 <  m(\pi^+\mu^+\mu^-)  < 2.4$ GeV$/c^2$.
The $D^+$ and $D^+_s$ signal yields are fixed based on the yields in the optimized $D^+_s$ sample and the relative efficiencies between
the $D_s^+$ and $D^+$ cuts.} \label{fig:dres2}
\end{figure}

\begin{figure}[h]
\centering
\includegraphics[width=80mm]{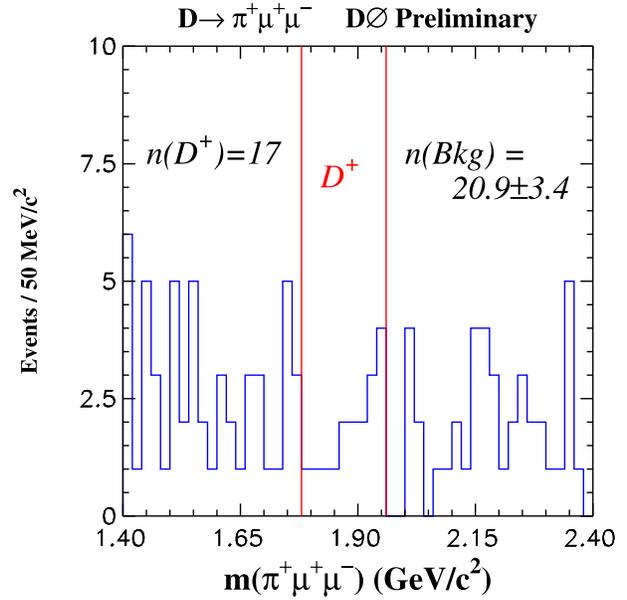}
\caption{Final search for $D^+ \rightarrow \pi^+ \mu^+\mu^-$  in the continuum region.}
\label{fig:dres3}
\end{figure}

The first goal is to observe the resonance production of 
$D_s\rightarrow \phi \pi \rightarrow \pi \mu^+\mu^-$.  The $D$ meson candidates for
 events where the dimuon system has an invariant mass consistent with a $\phi$ meson 
are shown in Fig~\ref{fig:dres1} for loose selection requirements and in Fig~\ref{fig:dres1a}
for requirements
 optimized for resonance production and background suppression. In both plots a $D_s$ signal is clearly observed
 and in the optimized selection the $D^+$ signal is also seen with a significance above 
background of $3.1$ sigma.  From this we extract the branching fraction 
$${\cal B}(D^+\rightarrow \phi\pi^+ \rightarrow \pi^+\mu^+\mu^-) = (1.75 \pm 0.7 \pm 0.5) \times 10^{-6} $$
which can be compared to the recent CLEO-c measurement of $(2.7^{+3.6}_{-1.8}\pm 0.2)\times 10^{-6}$ 
or the expected value of $1.77 \times 10^{-6}$ given by the product of the $D^+ \rightarrow \phi\pi^+$ 
and the $\phi\rightarrow \mu^+\mu^-$ branching fractions.

Having clearly established the resonance production, we turn to a search for non resonant $D$ production.
  The $D$ meson candidates for events passing tight requirements optimized for the non resonant analysis
 are shown in Fig~\ref{fig:dres2}.  The distribution on the left are again events where the dimuon
 system has an invariant mass consistent with a $\phi$ meson.  We expect to still see some excess from
 the $D_s$ and $D^+\rightarrow \phi\pi^+$ signals and indeed based on the yields from the above sample
 we predict $5.3 \pm 1.5$ events (signal plus background) in the $D^+$ signal region and $4.6 \pm 0.7$
 events in the $D_s$ signal region which compare well to the observed $6$ events in the $D^+$ window and $3$ events
 in the $D_s$ window.  This indicates that even after the tighter selection we still have sensitivity at
 the level of the $D^+$ branching fraction measured above. 

The distribution in Fig~\ref{fig:dres3} shows the $D$ candidates where the dimuon system has
 an invariant mass outside the $\phi$ window.  Here we see $17$ events in the $D^+$ window 
that are
 consistent with the expected background of $20.9 \pm 3.4$ events.  From this, we
 set the $90\%$ confidence level upper limit
$$ {\cal B}(D^+ \rightarrow \pi^+\mu^+\mu^-) < 4.7 \times 10^{-6}. $$ 
 This is the most stringent limit to date in the dimuon modes.
 Although this is approximately 
 500 times above the SM expected signal it is already below the allowed parameter space of SUSY R parity violating 
couplings.

\section{CP Violation in $B$ Mixing}

The CP eigenstates of neutral $B$ mesons are a linear combination of the flavor eigenstates
$$B_{CP} = pB^0 + q\bar{B}^0.$$
A difference in the coefficients $q \ne p$ would lead to CP violation $B$ mixing. 
 The signature of $B$ mixing is like-sign dilepton events
 indicating that the initially produced $B\bar{B}$ system decayed as either a $BB$
 or a $\bar{B}\bar{B}$ combination. The experimental signature for CP violation in $B$ mixing
 would be an asymmetry between the number of $l^+l^+$ and $l^-l^-$ final states.  

Since the current experimental limits are slightly below the percent level, our goal is to 
measure the charge asymmetry between like-sign dimuons at the tenth percent level precision.

At D\O\ , we can reach the desired accuracy by taking advantage of the two independent
 magnetic spectrometers: the inner solenoid and the outer muon system toroid. The field
 polarities are reversed frequently leading to roughly equal data sets with the four possible
 combinations of field configurations.  Comparing yields in these data sets allows us to make 
precise determinations of all detector induced asymmetries. 

The raw event yields for different combined toroid solenoid polarities are listed in Table~\ref{tab:nmumu} for our 
$1$ fb$^{-1}$ data set. The like-sign dimuon asymmetry after removing detector effects is 
$$A =  -0.0013 \pm 0.0012 \pm 0.0008.$$
To extract the underlying CP violating semileptonic asymmetry $A_{SL}$ from this number requires 
precise knowledge of the composition of our dimuon sample.  Besides our signal of direct 
semileptonic decay of both $b$ hadrons in the event, an important component of the sample
 is the coincidence of a primary muon from semileptonic b decay and a secondary muon 
from semileptonic charm decay.  Topologies like these that produce like-sign dimuons but
 are not associated with $B$ mixing dilute the asymmetry.

\begin{table}[h]
\begin{center}
\caption{Raw Results.}
\begin{tabular}{|l|c|c|}
\hline \textbf{Toroid*Solenoid Polarity} & \textbf{-1} & \textbf{+1}
\\
\hline N++ & 177,950 & 156,183  \\
\hline N-- -- & 176,939 & 156,148 \\
\hline N+ -- & 1,175,547 & 1,029,604 \\
\hline
\end{tabular}
\label{tab:nmumu}
\end{center}
\end{table}

The most problematic component are asymmetries induced by hadronic interactions of kaons
 in the detector.  Reactions like $K^- + N  \rightarrow \Lambda \pi$ have no $K^+$ analog.
  As the kaons travel through the detector the fraction of $K^+$ increases and  leaving a
 larger number of $\mu^-$ from $K^+$ decay in flight than $\mu^+$ from $K^-$ decay in flight.
We measure this asymmetry directly from reconstructed $B \rightarrow D^*\mu\nu$ events
 in the data sample and subtract it from the above asymmetry. 
 This is by far the largest source of systematic error in the analysis.

A sensitive cross check of the sample composition is the extraction of the integrated $B$
 mixing probability $\chi$ that can be extracted from the ratio of like-sign dimuons to
 opposite-sign dimuons.  In our data set we find
$$\chi = 0.136 \pm 0.001 \pm 0.024.$$
This can be compared to the Particle Data Group (PDG)~\cite{bib:PDF}  value of $\chi(\rm PDG) = 0.1281 \pm 0.0076$.
With our knowledge of the sample composition, we can then extract from the raw asymmetry
 the semileptonic asymmetry
$$A_{SL} =  -0.0044 \pm 0.0040 \pm 0.0028.$$
From which we extract the CP violation in $B$ mixing parameter $\epsilon_B$
$${{\cal R}(\epsilon_{B}) \over 1 + |\epsilon_B|^2} = -0.0011 \pm 0.0010 \pm 0.0007.$$
This measurement gives the best sensitivity to CP violation in $B$ mixing to date and
 can be compared to the world average value of all previous measurements
 $(0.0005 \pm 0.0031)$~\cite{bib:PDF}.

\section{$B_s$ Mixing}

Measurements of flavor mixing in neutral meson systems have historically led to profound insights into physics at
energy scales inaccessible to particle accelerators.   The discovery
of mixing and CP violation in the neutral kaon system~\cite{bib:kronin} led Kobayashi
and Maskawa to predict the existence of a third flavor generation~\cite{bib:KM}. The
discovery of mixing in the $B_d$ system gave the first indication that
the top quark is much heavier than the $W$
boson~\cite{bib:argus}. Finally, mixing induced CP
violation measurements in the $B_d$ system have precisely determined
the phase of the CP violating top quark coupling Arg$(V_{td}) \sim
24^\circ$~\cite{bib:belle}.  This motivates the study of mixing in the $B_s$ system.
A $B_s$ mixing measurement allows a precise determination of the magnitude of the CP
violating top coupling through the ratio $|V_{td}/V_{ts}|$ and may
lead to the discovery of new physics in $b\rightarrow s$ transitions~\cite{bib:tev-report}.

Measuring the $B_s$ oscillation frequency requires knowledge of the $B_s$ flavor at both
 production and decay.  Signatures of mixing are candidates where the flavor at decay is
 different than the flavor at production.  The mixing frequency can be determined by looking
 for an oscillation in the fraction of mixed events as a function of the $B_s$ decay length.

\begin{figure}[h]
\centering
\includegraphics[width=80mm]{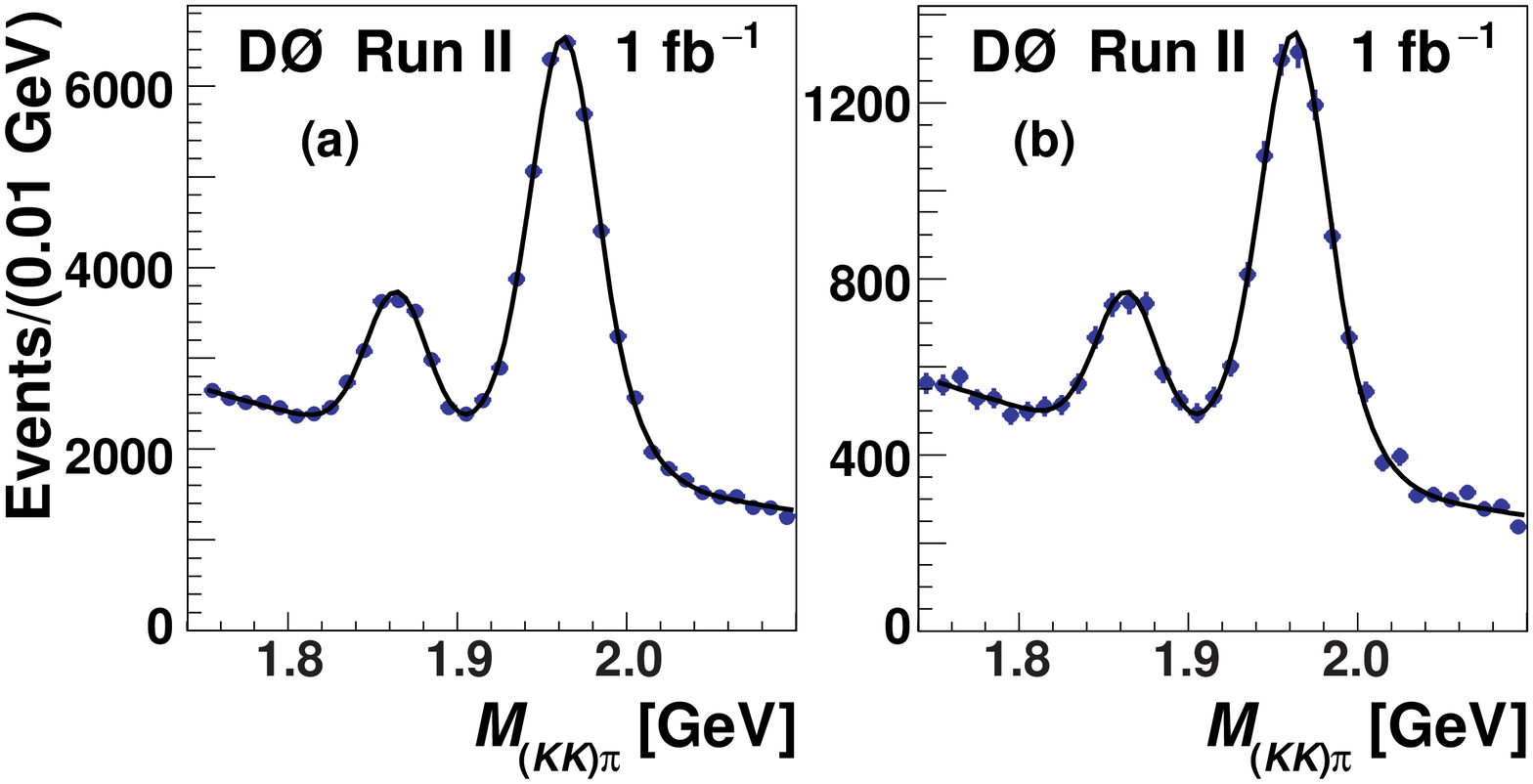}
\caption{$(K^+K^-)\pi^-$ invariant mass distribution for (a) the untagged $B^0_s$ sample, and (b) for 
candidates that have been flavor-tagged. The left and right peaks correspond to $\mu^+ D^-$ and $\mu^+ D^-_s$ candidates, respectively.
The curve is a result of fitting a signal plus background model to the data.  For fitting the mass spectra, a single Gaussian function
was used to describe the $D^- \rightarrow \phi \pi^-$ decays and a double Gaussian was used for the
$D^-_s \rightarrow \phi \pi^-$ decays.  The background was modeled by an exponential function.} \label{fig:mds}
\end{figure}

The $B_s$ flavor at decay is determined by reconstructing a flavor-specific final state. 
At D\O\ , we use the semileptonic decay $B^0_s \rightarrow D^-_s\mu^+\nu$, 
$D^-_s\rightarrow \phi\pi^-$, $\phi\rightarrow K^+K^-$ and charge conjugate states.
The $\phi\pi$ invariant mass distribution is shown in Fig~\ref{fig:mds}.  
In our 1 fb$^{-1}$ sample. we reconstruct approximately $26.7$ 
thousand $D_s\mu$ candidates predominantly from $B_s$ decay and $7.4$ thousand 
$D^+\mu$ candidates predominantly from $B^0$ decay.
The $B_s$ flavor at production is determined by examining inclusive properties
 of the system recoiling against the $B_s$ candidate.  In this way, we determine
 the flavor of the other $B$ hadron in the event that was pair produced with the
 $B_s$ candidate.  We use a likelihood based tagging algorithm based on momentum
 weighted sum of the charge of particles associated with either a lepton, a displaced
 vertex, or the entire recoil system.  Fig~\ref{fig:mds}b shows the event sample when the flavor tag is required.
The tagging purity of the combined likelihood is
 determined event by event and is calibrated by studying $B_d$ mixing in our 
$B^0_d\rightarrow D^{*-}\mu^+\nu$ sample.  The figure of merit for tagging is the 
effective efficiency, $\epsilon D^2$, that takes into account both the efficiency
 of the tagging algorithm and the weight that each event receives in the final
 analysis due to the purity of the tag.  In our $D^*\mu\nu$ sample, we measure an
 effective efficiency of 
$$\epsilon D^2 = 2.48 \pm 0.21^{+0.08} _{-0.06}.$$

The decay length of the $B_s$ meson is taken as the transverse distance from the
  primary vertex reconstructed using tracks originating in the beam spot, and the
 secondary vertex reconstructed from the $D_s$ and $\mu$ particles.  To boost the
 decay length to the proper frame, we use MC to simulate the missing energy from
 the neutrino or other pions or kaons from $B_s\rightarrow D_sX$ decays. 

We require
 a very accurate determination of our decay length resolution so that we can
 distinguish a lack of a signal in our data from a lack of the ability to resolve a 
frequency beyond our resolution. We test our resolution model using prompt $J/\psi$ 
candidates from $p\bar{p} \rightarrow J/\psi X$ interactions.  Here the $J/\psi$ decays 
at the primary vertex so that our measurement of the distance from the primary vertex 
to the $J/\psi$ vertex is a direct measurement of our resolution.

\begin{figure}[h]
\centering
\includegraphics[width=80mm]{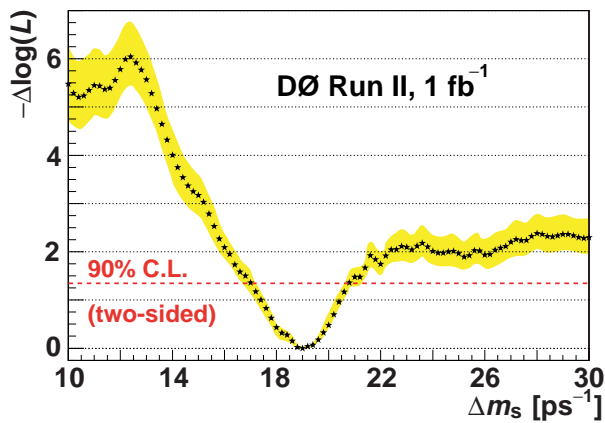}
\caption{Value of $-\Delta log \mathcal {L}$ as a function of $\Delta m_{s}$. Star symbols do not include systematic uncertainties, and 
the shaded band represents the envelope of all $log \mathcal {L} $ scan curves due to different systematic uncertainties.} \label{fig:bsmix2}
\end{figure}

Fig~\ref{fig:bsmix2} shows the value of $-\Delta log \mathcal {L}$ as a function 
of $\Delta m_{s}$, indicating a favored value of $19 ps^{-1}$, while variations of 
-$log \mathcal {L} $ from the minimum indicates an oscillation frequency of $17 < \Delta m_s <21 ps^{-1}$
at the 90\% C.L. The uncertainties are approximately Gaussian inside the interval.
Using 1000 parameterized Monte Carlo samples, it was determined that for a true value of 
$\Delta m_s = 19 ps ^{-1}$, the probability was 15\% for measuring a value in the range  
$17 < \Delta m_s <21 ps^{-1}$ with a $-\Delta log \mathcal {L}$ lower by at least 1.9 than the corresponding value at 
$\Delta m_s = 25 ps ^{-1}$. 

\begin{figure}[h]
\centering
\includegraphics[width=80mm]{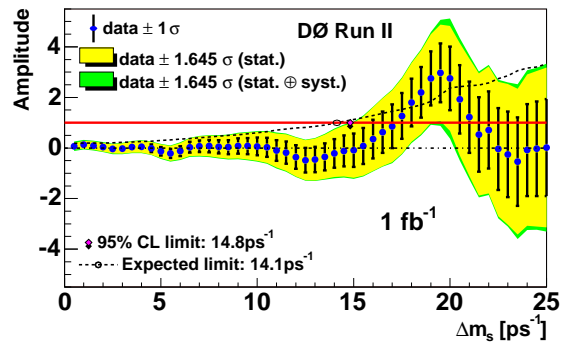}
\caption{$B^0_s$ oscillation amplitude as a function of oscillation frequency, $\Delta m_s$.} \label{fig:bsmix3}
\end{figure}

The amplitude method~\cite{bib:ampl} was also used and the results are shown in Fig~\ref{fig:bsmix3}.
The unbinned likelihood fit is repeated including the oscillation amplitude $\mathcal{A}$
as a function of the input value of $\Delta m_{s}$. At $\Delta m_{s}= 19 ps^{-1}$ the measured data point deviates from the hypothesis 
$\mathcal{A}= 0$ ($\mathcal{A}= 1$) by 2.5 (1.6) standard deviations, corresponding to a two-sided C.L. of 1\% (10\%), 
and is in agreement with the likelihood results.  
This is the first report of a direct two-sided bound on the $B^0_s$ oscillation frequency.

\section{Conclusions}
D\O\ preliminary results based on approximately $1$ fb$^{-1}$ of $p \bar{p}$
collisions at $\sqrt{s} = 1.96$ TeV recorded at the Fermilab Tevatron have been presented.
We set the most stringent limit to date in a search for the flavor changing neutral current in the 
dimuon modes.  At the $90\%$ confidence level upper limit we find
$ {\cal B}(D^+ \rightarrow \pi^+\mu^+\mu^-) < 4.7 \times 10^{-6}. $ 
We have extracted the CP violation in $B$ mixing parameter $\epsilon_B$
$${{\cal R}(\epsilon_{B}) \over 1 + |\epsilon_B|^2} = -0.0011 \pm 0.0010 \pm 0.0007,$$
which is the best sensitivity to CP violation in $B$ mixing to date.
We have presented the first report of a direct two-sided bound on the $B^0_s$ oscillation frequency,
$17 < \Delta m_s <21 ps^{-1}$ at the 90\% C.L.


\bigskip 

\end{document}